\documentclass[aps,prl,twocolumn,superscriptaddress,groupedaddress]{revtex4}
\pdfoutput=1
\usepackage[version=3]{mhchem} 
\usepackage[dvips]{epsfig}
\usepackage{graphicx}  
\usepackage{dcolumn}   
\usepackage{bm}        
\usepackage{amssymb}   
\usepackage{color}

\begin{document}
\author{Matthew Mitchell}
\affiliation{Institute for Quantum Science and Technology, University of Calgary, Calgary, AB, T2N 1N4, Canada}
\affiliation{National Institute for Nanotechnology, 11421 Saskatchewan Dr.\ NW,  Edmonton, AB T6G 2M9, Canada}
\author{Aaron C. Hryciw}
\affiliation{National Institute for Nanotechnology, 11421 Saskatchewan Dr.\ NW, Edmonton, AB T6G 2M9, Canada}
\author{Paul E. Barclay}
\email{pbarclay@ucalgary.ca}
\affiliation{Institute for Quantum Science and Technology, University of Calgary, Calgary, AB, T2N 1N4, Canada}
\affiliation{National Institute for Nanotechnology, 11421 Saskatchewan Dr.\ NW,  Edmonton, AB T6G 2M9, Canada}

\title{Cavity optomechanics in gallium phosphide microdisks}

\begin{abstract}
We demonstrate gallium phosphide (GaP) microdisk optical cavities with intrinsic quality factors $ > 2.8\times10^{5}$ and mode volumes $< 10 (\lambda/n)^3$, and study their nonlinear and optomechanical properties. For optical intensities up to $8.0\times10^4$ intracavity photons, we observe optical loss in the microcavity to decrease with increasing intensity, indicating that saturable absorption sites are present in the GaP material, and that two-photon absorption is not significant.  We observe optomechanical coupling between optical modes of the microdisk around 1.5 $\mu$m and several mechanical resonances, and measure an optical spring effect consistent with a theoretically predicted optomechanical coupling rate $g_0/2\pi \sim 30$ kHz for the fundamental mechanical radial breathing mode at 488 MHz.
 \end{abstract}

\maketitle

\noindent

Cavity optomechanics provides a platform for exquisitely controlling coherent interactions between photons and mesoscopic mechanical excitations \cite{ref:kippenberg2007com, ref:aspelmeyer2013co}.  Nanophotonic implementations of cavity optomechanics have recently been used to demonstrate laser cooling\cite{ref:park2009rsc, ref:schliesser2009rsc, ref:chan2011lcn, ref:safavinaeini2012oqm}, optomechanically induced transparency\cite{ref:weis2010oit, ref:safavi2011eit}, and coherent wavelength conversion \cite{ref:hill2012cow, ref:dong2012odm, ref:liu2013eit}.  These experiments were  enabled by  photonic micro- and nanocavities engineered to minimize optical and mechanical dissipation rates, $\gamma_\text{o}$ and $\gamma_\text{m}$ respectively, while enhancing the single-photon optomechanical coupling rate, $g_0$. The degree of coherent coupling between photons and phonons in these devices is often described by the cooperativity parameter, $C = N g_0^2 / \gamma_\text{o}\gamma_\text{m}$, which may exceed unity in several cavity optomechanics systems under investigation\cite{ref:aspelmeyer2013co} for a sufficiently large intracavity photon number, $N$.  Of particular promise are nanophotonic cavity optomechanical systems realized from semiconductors, such as silicon optomechanical crystals \cite{ref:eichenfield2009oc, ref:chan2011lcn} and gallium arsenide microdisks \cite{ref:ding2010hfg}, owing to the tight optical confinement and large $g_0$ possible in these high refractive index contrast structures. Increasing $C$ (above, for example, the value of $C \sim 20$ demonstrated in optomechanical crystals \cite{ref:chan2012ooc}), would enable improved bandwidth of coherent wavelength conversion \cite{ref:hill2012cow}, observation of normal mode splitting \cite{ref:groblacher2009osc}, and faster optomechanical cooling \cite{ref:chan2011lcn}.  However, in these semiconductor-based large-$g_0$ devices, $N$ is clamped by two-photon absorption\cite{ref:barclay2005nrs, ref:sun2013noe}.  Here we demonstrate a cavity optomechanical system realized in a high refractive index material which does not exhibit nonlinear optical loss at commonly used telecommunications wavelengths.

\begin{figure}
\begin{center}
\epsfig{figure=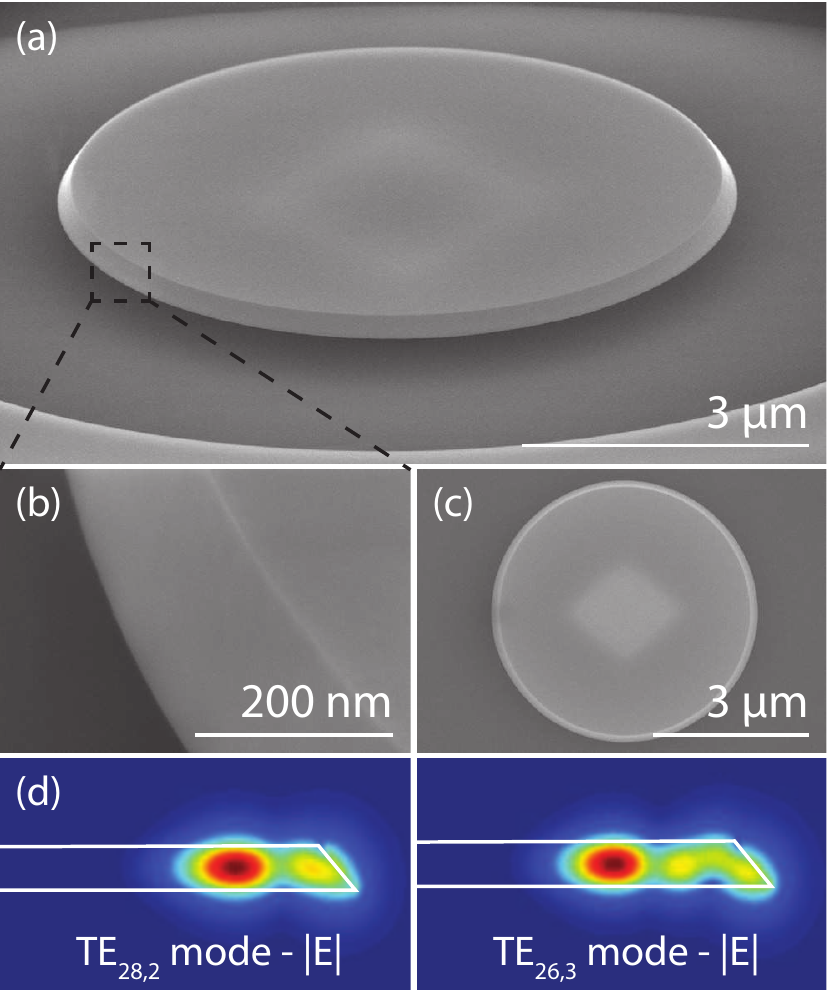, width=1\linewidth}
 \caption{(a)--(c) Scanning electron micrographs of a GaP microdisk resonator with a close-up of the disk sidewall. (d) Electric field distributions of the second ($n=2$) and third ($n=3$) radial order TE$_{m,n}$ microdisk modes, where $m$ represents the azimuthal mode index and is chosen such that the mode wavelength is near $1550$ nm.}
\label{fig:schematic}
\end{center}
\end{figure}

The optomechanical system studied in this letter, an example of which is shown in Fig.\ \ref{fig:schematic}(a)--(c), is based on whispering-gallery mode microdisk optical cavities fabricated from gallium phosphide (GaP). GaP is a semiconductor with a desirable combination of high refractive index ($n_\text{GaP} \sim 3.05$ at 1550 nm) and large electronic bandgap (2.26 eV); it is therefore optically transparent from 550 nm to IR wavelengths and has low two-photon absorption at 1550 nm.  Compared to other materials with similar transparency windows used in cavity optomechanics, such as SiO$_2$ ($n_{\text{SiO}_2}= 1.45$) \cite{ref:schliesser2009rsc, ref:weis2010oit, ref:park2009rsc, ref:dong2012odm}, Si$_3$N$_4$ ($n_\text{Si$_3$N$_4$}$ = 2.0--2.2)  \cite{ref:eichenfield2009apn, ref:liu2013eit} and AlN ($n_\text{AlN}$ = 2.0--2.2) \cite{ref:xiong2012ihf},  GaP has a larger refractive index, enabling the fabrication of optical nanocavities with ultrasmall mode volumes\cite{ref:rivoire2008gpp, ref:barclay2011hnr}.  As such, GaP is a promising material for realizing cavity optomechanical systems with $g_0/2\pi$ approaching 1 MHz, as observed in Si and GaAs devices \cite{ref:eichenfield2009oc, ref:chan2012ooc, ref:ding2011wsg, ref:ding2010hfg, ref:sun2012fdc}.  In this letter, we show that GaP microcavities combine high optical quality factor $Q_i \sim 2.8 \times 10^5$, large optomechanical coupling $g_0/2\pi > 30$ kHz, and no observed two-photon absorption at 1550 nm band wavelengths. These properties allow the devices to operate close to the resolved-sideband regime with an intracavity photon number sufficient to allow observation of the optical spring effect.

The microdisks studied here were fabricated from an epitaxially grown wafer (supplied by IQE) consisting of a 250-nm-thick GaP device layer supported by a 750-nm-thick sacrificial aluminum gallium phosphide (AlGaP) layer and a GaP substrate. Microdisk patterns were defined using electron-beam lithography with ZEP520A resist, followed by a resist reflow step at $165~^{\circ}{\rm C}$ for 5 min.  This step decreases imperfections in the circular electron-beam resist pattern, resulting in smoother sidewalls and an increased $Q_i$ \cite{ref:borselli2005brs}. The resulting pattern was transferred into the GaP layer using a low-DC-bias inductively-coupled plasma reactive-ion etch with Ar/Cl$_2$ chemistry.  The reflowed nature of the mask results in an angled GaP sidewall etch profile during this step, as seen in Figs.\ 1(a)--(c). This angle is not expected to limit $Q_i$ in these devices, as microdisks with similar profiles \cite{ref:borselli2005brs,ref:barclay2006ifc} have been reported with $Q_i > 10^6$ in other material systems.  The resist was then removed with a 10 min deep-UV exposure (1.24 mW/$\text{cm}^2$ at 254 nm), followed by a 5 min soak in Remover PG$\circledR$, and subsequent rinsing in acetone, isopropyl alcohol, and de-ionized water. The microdisks were undercut by selectively removing the AlGaP layer using a hydrofluoric acid wet etch (H$_2$O: 49\% HF = 3:1). Microdisks with radii of $\sim$ 4 $\mu$m were studied in the work presented below.

 \begin{figure}
\begin{center}
\epsfig{figure=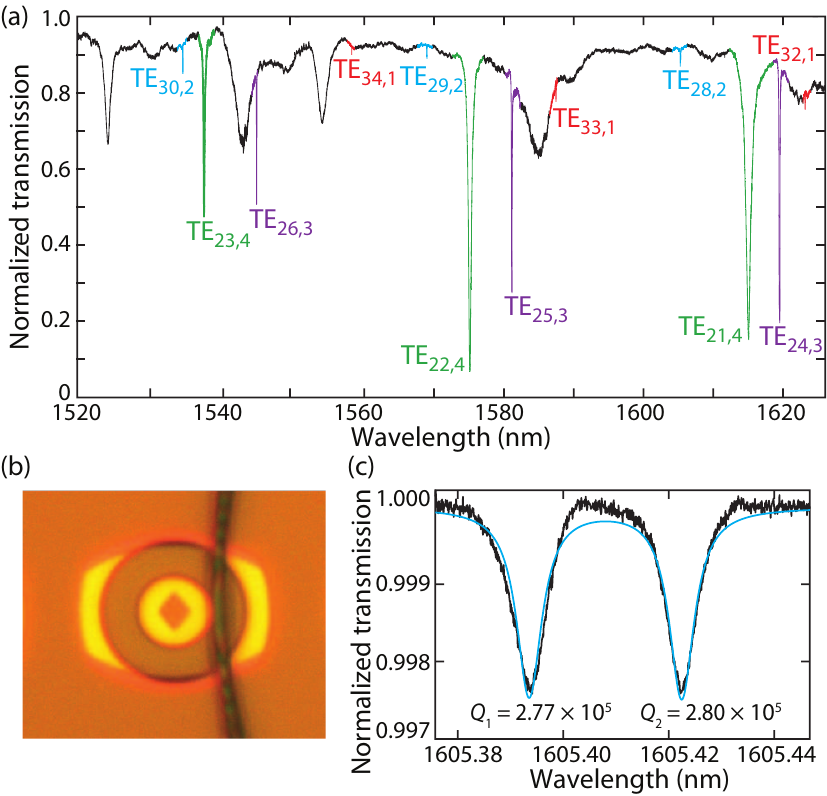, width=1\linewidth}
 \caption{(a) Fiber taper transmission spectrum of a GaP microdisk with a radius of 4 $\mu$m, normalized by the fiber taper transmission when it is positioned far from the microdisk. Modes with common free spectral range are color coded and labeled as TE$_{m,n}$, with azimuthal and radial mode indices $m$ and $n$, respectively. (b) Microscope image of dimpled fiber taper side-coupled to microdisk. (c) Narrow-wavelength scan of a high-$Q_i$ mode, associated with the $\left[m,n\right]$ = 28,2 mode indices.}
\label{fig:optical}
\end{center}
\end{figure}

The optomechanical properties of the microdisks were probed using a dimpled optical fiber taper \cite{ref:michael2007oft} to evanescently couple light into and out of the cavity optical modes. The dimpled fiber taper was fabricated by modifying the process of Michael \emph{et al.}\ \cite{ref:michael2007oft} to use a ceramic edge as the dimple mold.  The resulting dimple can be positioned within the optical near field  of the microdisk, as shown in Fig.\ \ref{fig:optical}(b).  Two tunable laser sources (New Focus Velocity) were used to measure the transmission of the fiber taper over the 1520--1625 nm wavelength range.  The transmitted signal was split using a 10:90 fiber coupler, detected using low- (Newport 1621) and high-speed (Newport 1554-B) photodetectors, and recorded using a data acquisition card and a real-time spectrum analyser (Tektronix RSA5106A), respectively.

\begin{figure}
\begin{center}
\epsfig{figure=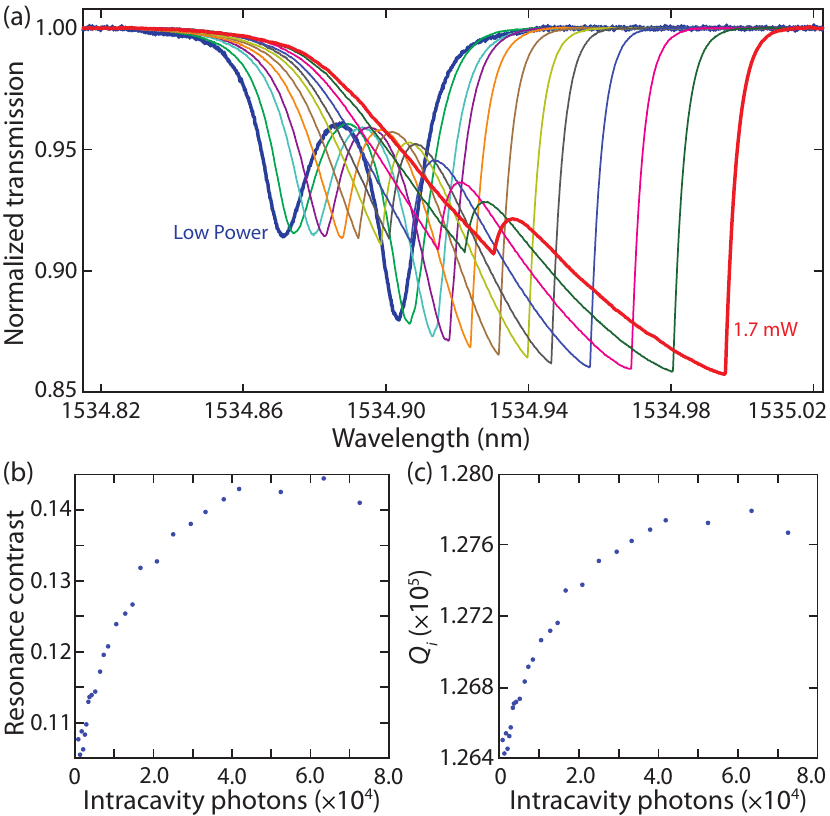, width=1\linewidth}
 \caption{(a) Optical response of the TE$_{30,2}$ mode,  measured for varying input power $P_i$. (b) Measured resonance contrast $\Delta T_o$, and (c) increase of intrinsic optical quality factor $Q_i$, as a function of intracavity photon number $N$, for the TE$_{34,1}$ mode.}
\label{fig:nonlinear}
\end{center}
\end{figure}

The optical response of a typical microdisk as a function of input wavelength, shown in Fig.\ \ref{fig:optical}(a) when the fiber taper dimple is positioned in the microdisk near field as in \ref{fig:optical}(b), consists of numerous resonances corresponding to excitation of high-$Q_i$ whispering-gallery modes (WGMs) of the microdisk. The modal indices of each resonance, describing their azimuthal ($m$) and radial ($n$) order, were determined from comparisons of the measured $Q_i$ and mode spacing with the mode spectrum predicted by finite-difference time-domain (FDTD) simulations \cite{ref:oskooi2010mff} of the device. Note that this identification of azimuthal mode number should be considered approximate as our simulations used the nominal refractive index of GaP, and do not take into account perturbations to the refractive index from material impurities. For the device geometry used here, only TE-like modes (electric field dominantly radially polarized) with fundamental vertical field profile have sufficiently high $Q_i$ to be observed.  Figure \ref{fig:optical}(c) displays the optical response of the highest-$Q_i$ mode observed in the fabricated GaP microdisks, which has $Q_i \sim 2.8 \times 10^5$ with minimal fiber loading. These are the highest optical quality factors  observed to date in GaP nanophotonic devices. A mechanism limiting $Q_i < 10^6$ in these disks is the pedestal height (750 nm), which is determined by the thickness of the AlGaP layer. The choice of AlGaP layer thickness was determined by material availability, and should be further optimized in future devices. The doublet resonance structure is a result of scattering and modal coupling from imperfections in the microdisk, and indicates that the observed modes are standing waves \cite{ref:borselli2004rsm}. Based on the comparisons with simulations, this mode is predicted to have modal indices $\left[m,n\right]=\left[28,2\right]$, field profile shown in Fig.\ \ref{fig:schematic}(d), and standing-wave mode volume $V_\text{o} =  9.7 (\frac{\lambda}{n_\text{GaP}})^3$.

To probe for nonlinear optical absorption in these microdisks, we measured their response as a function of fiber taper input power, $P_i$.  Figure \ref{fig:nonlinear}(a) shows the fiber taper transmission for varying  $P_i$ when the source laser is scanned with increasing wavelength across the TE$_{30,2}$ microdisk resonance. Although a power-dependent optical response is clearly observed, its behavior is indicative of a decrease in the internal optical loss of the microdisk, $\gamma_i$, with increasing $P_i$.  This anomalous effect is illustrated in  Fig.~\ref{fig:nonlinear}(b) for the TE$_{34,1}$ mode, which shows that the resonance contrast, $\Delta T_o = 4K/(1+K)^2$, increases with $P_i$, where $K = \gamma_e /(\gamma_e+\gamma_i(P_i))$, and $\gamma_e$ is coupling rate between the microdisk standing wave mode and the forward- or backward-propagating fiber taper mode (i.e., $\gamma_\text{o} = \gamma_i + 2\gamma_e$).  This is contrary to the behavior of cavities exhibiting multiphoton absorption, such as those fabricated from Si \cite{ref:barclay2005nrs}. Figure \ref{fig:nonlinear}(c) shows the corresponding increase of $Q_i$ as a function of $N$, calculated according to:
  \begin{equation}
  \hbar\omega_\text{cav}N = \Delta T_o P_i\frac{Q_i}{\omega_\text{cav}}.
  \end{equation}
This effect may be the result of saturable absorbers in the microcavity, possibly in the form of O$_2$-related impurities \cite{ref:henry1968ecl,ref:dean1968ida} and will be investigated further in future work. A redshift in the cavity resonance wavelength $\lambda_\text{o}$, is also observed with increasing $P_i$.  This dispersive effect is the result of heating of the microcavity due to optical absorption, and can be suppressed by operating at lower temperatures \cite{ref:liu2013eit}.

 \begin{figure}
\begin{center}
\epsfig{figure=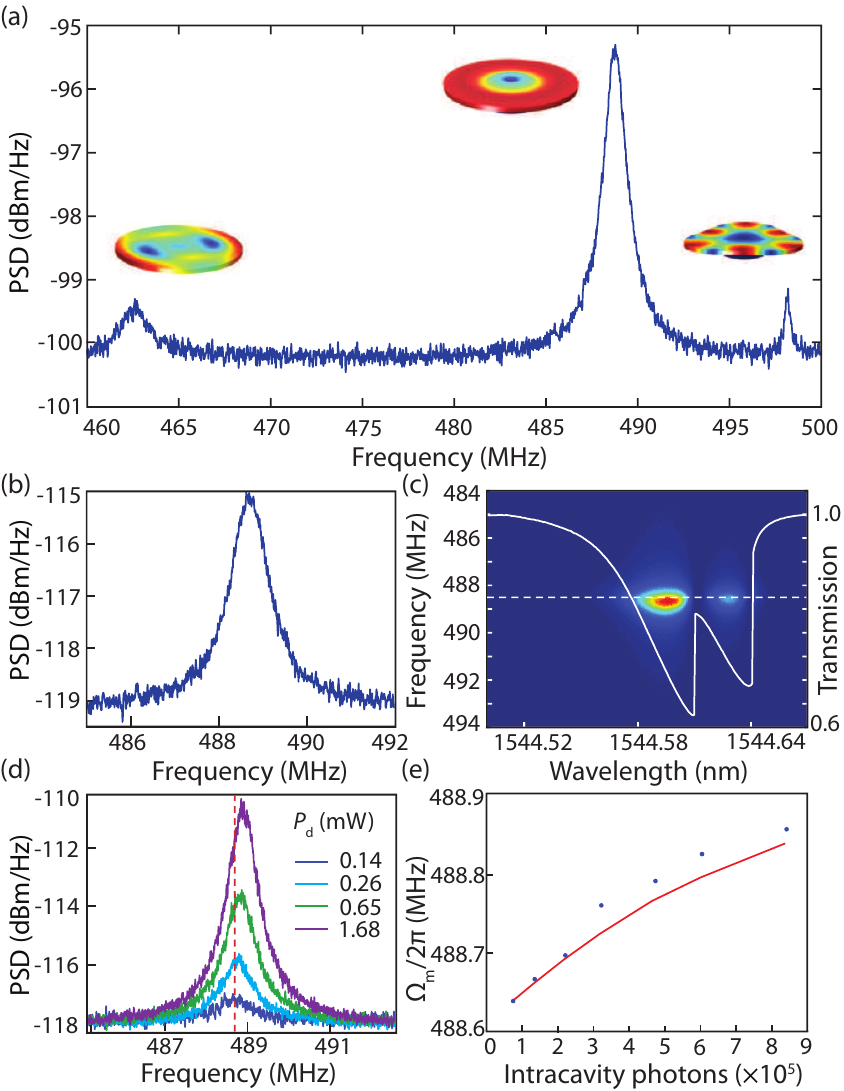, width=1\linewidth}
 \caption{Optomechanical data for a 4 $\mu$m radius GaP microdisk. (a) Electronic power spectral density (PSD) of photodetected taper transmission when $\omega_\ell$ is blue detuned by $\sim \gamma_\text{o}/2$ from the TE$_{26,3}$ optical resonance. Simulated displacement amplitude profiles of the mechanical modes corresponding to the observed resonances in the PSD are also shown. (b) PSD of the fundamental RBM, measured for $P_i \sim 100~\mu$W. (c) Mechanical response of the RBM for varying laser wavelength for $P_i \sim 2.1$ mW.  The corresponding transmission spectrum of the fiber taper evanescently coupled to the microdisk (white line), and low power RBM frequency (white dashed line), are also shown. (d) PSD of the fundamental RBM at four optical dropped powers when the laser was blue detuned by $\sim \gamma_\text{o}/2$ from $\omega_\text{c}$. (e) Resonance frequency of the RBM as a function of intracavity photon number, when the laser was blue detuned by $\sim \gamma_\text{o}/2$ from $\omega_\text{c}$. The solid red line represents predicted values from a theoretical model (see text).}
\label{fig:optomechanics}
\end{center}
\end{figure}

To study the optomechanical properties of the microdisks, the effect of the thermal motion of the microdisk mechanical modes on the noise spectrum of the optical power transmitted through the fiber taper was measured. The dimensions of the microdisks were chosen to maximize $g_0$ and the mechanical frequency, $\Omega_\text{m}$, of the fundamental radial breathing mode (RBM), without compromising $Q_i$. The 4 $\mu$m radius of the microdisks studied here is small enough to allow the mechanical breathing mode to have a high frequency ($\Omega_\text{m}/2\pi\sim$ 0.5 GHz) and large enough to avoid optical radiation loss at a rate above other loss mechanisms of the microdisk. A typical spectrum, obtained in ambient conditions using the TE$_{26,3}$ mode, with input laser frequency $\omega_\ell$ blue detuned ($\omega_\ell \approx \omega_\text{c} + \gamma_\text{o}/2$), where $\omega_\text{c}$ is the cold cavity optical resonance frequency, is given in Fig.~\ref{fig:optomechanics}(a). Peaks associated with three mechanical resonances are visible; of particular interest is the fundamental RBM at $\Omega_\text{m}/2\pi = 488$ MHz. The RBM was found to have a mechanical zero-point fluctuation amplitude of 0.66 fm and displacement sensitivity of $4.43\times10^{-17} \text{m}/\sqrt{\text{Hz}}$ for $P_i$ of $\sim$ 600 $\mu$W. The mode profiles of the mechanical resonances, and their effective mass\cite{ref:aspelmeyer2013co}, $m_\text{eff}$, were identified using  COMSOL finite element simulations. The fundamental RBM ($m_\text{eff}=40$ pg) was found to have a mechanical quality factor, measured at low $P_i$ to avoid optical backaction effects, of $Q_\text{m} \sim$ 640, while the $\Omega_\text{m}/2\pi = 462$ MHz and $\Omega_\text{m}/2\pi = 498$ MHz modes have $Q_\text{m} \sim$ 360 and 450, respectively.  $Q_\text{m}$ can be improved by engineering the device geometry and testing environment. While the RBM should not be limited by squeeze film damping, operation in vacuum could increase $Q_\text{m}$ by eliminating air damping \cite{ref:riviere:2011cos,ref:tang2012hqs,ref:hiebert2012ndf, ref:parrain2012dod}. Mechanical clamping and phonon radiation loss can be reduced by further undercutting the microdisks to create ultra-small pedestals \cite{ref:liu2013eit, ref:schliesser2008rsc, ref:nguyen2013uqf}. Phononic Bragg mirror inspired pedestal structures as seen in Ref.\cite{ref:nguyen2013uqf} could also be used to reduce clamping losses. This would require the growth of hetero-structured wafers, which would also allow control over pedestal height, as well as provide an opportunity to reduce impurities present in the current device layer. Through a combination of these techniques a $10^2$ increase in $Q_\text{m}$ is reasonable.

The optomechanical coupling coefficient $G$ and rate $g_0$\cite{ref:eichenfield2009oc,ref:aspelmeyer2013co} between the fundamental RBM and the various optical WGMs of the microdisk were calculated using COMSOL finite element simulations and are given in Table I. For the devices under study, $g_0/2\pi \sim 30$ kHz and $G/2\pi \sim 50$ GHz/nm are comparable to record values\cite{ref:ding2010hfg}.  The realizable cooperativity parameter for this device can be predicted from $g_0$ and the measured dissipation rates, $\gamma_\text{m}/2\pi \sim 7.63\times10^{5}$ Hz and  $\gamma_\text{o}/2\pi \sim 6.68\times10^{8}$ Hz, of the high-$Q_i$ TE$_{28,2}$  mode. These rates correspond to a single-photon cooperativity of $\sim 1.3\times10^{-6}$, which is on the same order of magnitude as similar WGM microcavity devices \cite{ref:park2009rsc,ref:schliesser2009rsc,ref:aspelmeyer2013co}. It is predicted that $C >$ 1 for $N >$ $7.53\times10^{5}$ intracavity photons for the TE$_{28,2}$ mode. The largest experimental $C$ inferred was $C\sim 0.53$ for the TE$_{30,2}$ mode, with $N \sim 4.15\times 10^{5}$ (corresponding to $\sim$ 700 $\mu$W of dropped power). Note that this $C$ is not a fundamental limit of our system, as $C$ can be enhanced in future by improving the fiber--microdisk coupling efficiency or increasing $P_i$.

The measurement noise background in Fig.~\ref{fig:optomechanics} was dominated by detector noise ($\sqrt{S^{DET}_{xx}}\sim ~ 2 \times 10^{-17}$ m/$\sqrt{\text{Hz}}$), and shot noise ($\sqrt{S^{SN}_{xx}}\sim ~ 5 \times 10^{-18}$ m/$\sqrt{\text{Hz}}$). This noise level is roughly a factor of 100 times the standard quantum limited displacement noise ($\sqrt{S^{SQL}_{xx}} = 6 \times 10^{-19}$ m/$\sqrt{\text{Hz}}$). For the current device, at higher operating $P_i$, where optomechanical backaction noise is equal to shot noise ($S^{BA}_{xx} = S^{SN}_{xx}$), detector noise will remain dominant. As such, detection of the RBM with SQL precision using the TE$_{26,3}$ optical mode, would require lower noise photodetection. Alternatively, using higher $Q_i$ modes such as the TE$_{28,2}$ mode in Fig.\ \ref{fig:optical}(c), would further increase the measurement precision. However, reaching the SQL using these high-$Q_i$ modes will not be possible without significantly improved fiber coupling efficiency, as critical coupling is a requirement for SQL measurements \cite{ref:aspelmeyer2013co}.

\begin{table}
\caption{\label{tab:example} Simulated (FDTD) radiation-loss-limited optical quality factor for substrate-free  and substrate-limited cases ($Q^{\text{free}}_{\text{rad}}$ and $Q^{\text{sub}}_{\text{rad}}$, respectively), measured intrinsic optical quality factors ($Q_i$), and simulated optomechanical coupling parameters for fundamental RBM ($\Omega_\text{m}/2\pi$ = 488 MHz), for select measured optical modes.}
\begin{ruledtabular}
\begin{tabular}{cccccc}

Mode & $Q^{\text{free}}_{\text{rad}}$ & $Q^{\text{sub}}_{\text{rad}}$ & $Q_i$ & $G/2\pi$ ($\frac{\text{GHz}}{\text{nm}}$) & $g_0/2\pi$ (kHz)\\[1mm]
\hline
TE$_{33,1}$ & $ > 10^8$ & 1.1$\times10^6$ & 2.6$\times10^5$& 48 & 31 \\
TE$_{28,2}$ & $ > 10^8$ & 3.9$\times10^5$ & 2.8$\times10^5$& 40 & 26 \\
TE$_{25,3}$ & 6.0$\times10^7$ & 1.4 $\times10^5$ & 0.9$\times10^5$& 35 & 23 \\
TE$_{22,4}$ & 1.8$\times10^5$  & 2.1$\times10^4$ & 0.5$\times10^4$& 32 & 21 \\

\end{tabular}
\end{ruledtabular}
\end{table}

We further explored the optomechanical coupling by measuring the dependence of the lineshape of the fundamental RBM on $N$. Figure \ref{fig:optomechanics}(c) illustrates the dependence of the mechanical spectrum on laser wavelength as it is swept across the TE$_{26,3}$ optical cavity resonance, at a fixed $P_i = 2.1$ mW. As the laser approaches the cavity resonance and $N$ increases, $\Omega_\text{m}$ is observed to increase by up to 250 kHz. This is the result of the optical spring effect, and is further illustrated in Figs.\ \ref{fig:optomechanics}(d) and (e), which show the effect of varying $P_i$ for fixed $\omega_\ell = \omega_\text{c} + \gamma_\text{o}/2$. Note that for $P_i \gg 200\mu$W, $\omega_\text{c}$ red-shifts due to the heating of the microdisk by the intracavity saturable absorption described above.  The fit in Fig.\ \ref{fig:optomechanics}(e) was obtained using the model for the optical spring effect described by Aspelmeyer \emph{et al.}\ \cite{ref:aspelmeyer2013co}, taking into account the measured dependence of $\Delta\omega = \omega_\ell - \omega_\text{c}$, and $\Delta T_o$ on $N$. The observed power dependent shift in $\Omega_\text{m}$ of up to 218 kHz is in good agreement with this model. We also observe mechanical damping while blue detuned from the cavity resonance, as can be observed from Fig.\ \ref{fig:optomechanics}(d). This effect is a result of thermo-optic damping\cite{ref:eichenfield2009apn,ref:srinivasan2011oti}. For small $N$, the TE$_{26,3}$ mode used in these spring effect measurements had an intrinsic $Q_{i}  \sim 6.3 \times 10^4$ and a resonance contrast of $\Delta T_o \sim 0.63$. From FDTD simulations, this mode is predicted to have a standing-wave mode volume of $V_\text{o} =  11.1 (\frac{\lambda}{n_\text{GaP}})^3$ and the field profile shown in Fig.\ \ref{fig:schematic}(d).

For optomechanical cooling of a mechanical resonator, it is desirable to be in the sideband-resolved regime \cite{ref:schliesser2008rsc,ref:park2009rsc}. In our case, the optical linewidth of the highest-$Q_i$ mode is $\gamma_o/2\pi \sim 668$ MHz. As seen in Table I, our measured $Q_i$ values are significantly lower than the radiation-loss-limited quality factor, $Q^{\text{sub}}_{\text{rad}}$, which includes leakage into the substrate. We also observe that the TE$_{33,1}$ and TE$_{28,2}$  modes have similar $Q_i$, despite the lower predicted radiation loss of the TE$_{33,1}$ mode.  This suggests that either material absorption or surface scattering is placing the upper limit on $Q_i$ for these devices.  The sideband-resolved regime may be attainable through further optimization of $Q_i$ by fabricating devices from wafers with a thicker sacrificial AlGaP layer and lower GaP linear absorption, and with reduced surface roughness.

In conclusion, we have fabricated GaP microdisks with the highest intrinsic optical quality factors to date ($Q_i > 2.8 \times 10^{5}$) and have demonstrated the first instance of cavity optomechanics in GaP devices. The observed increase in $Q_i$ with input power, paired with the absence of nonlinear absorption for $N > 10^5$, illustrates the potential that GaP holds for experiments in quantum optomechanics.  These devices are also promising for implementing high-frequency cavity optomechanics at visible wavelengths and for applications in nonlinear optics.

We would like to thank Marcelo Wu, David Lake, Behzad Khanaliloo, and Chris Healey for helpful discussions. We would also like to express gratitude to Charles Santori, Kai-Mei C.\ Fu, and Hewlett-Packard Labs for valuable contributions providing and characterizing the GaP material. This work was supported by NRC, CFI, iCORE/AITF, and NSERC.

\bibliographystyle{osa} \bibliography{nano_bib}

\end{document}